\documentstyle[preprint,aps]{revtex}
\begin{document}
\draft
\preprint{ }
\title {Equilibrium current and orbital magnetization in a quantum Hall
fluid}
\author{ Michael R. Geller and G. Vignale} \address{Department of Physics,
University of Missouri, Columbia, Missouri 65211}
\date{\today}
\maketitle

\begin{abstract}
We present a general theory for the equilibrium current distribution in an
interacting two-dimensional electron gas subjected to a perpendicular
magnetic field, and confined by a potential that varies slowly on the scale of
the magnetic length.  The  distribution is found to consist of strips or
channels of current, which alternate in direction,  and which have universal
integrated strength.
\end{abstract}
\pacs{  }
\narrowtext

{\bf 1. Introduction}

The quantum Hall effect \cite{Klitzing} is  described  as the quantization of
the off-diagonal {\it conductance} of a two-dimensional electronic device, at
integral or simple fractional multiples  of $e^2 /h$, and the vanishing of the
diagonal conductance,   when a perpendicular magnetic field of appropriate
strength is present.   It has long been understood \cite{Hall} that this
``appropriate strength" of the magnetic field is such that the the Fermi level
lies in a {\it gap} of the spectrum of bulk extended states.  Finer features
of the effect are  not nearly as well understood.  There is
considerable confusion, for instance, concerning  the question of the physical
distribution of the Hall current in a quantum Hall device. Is the Hall current
carried by ``edge states" residing near the the edges of the device, or is
it carried by the bulk of the electron fluid?  While early work
\cite{Halperin},  based on a model of non-interacting electrons, supported the
first alternative, more recent work \cite{Thouless} taking into account
Coulomb interactions self-consistently, concludes that at least a
substantial part of the current is carried in the bulk. Experimental attempts
to resolve the ambiguity by directly imaging the voltage  and/or
current distribution, have been made \cite {Diener},  with no conclusive
results so far. A major practical difficulty is  that in a nonuniform
electron fluid subjected to a magnetic field, a finite current density exists
even in thermal equilibrium,  due to the time-reversal symmetry  breaking
action
of the magnetic field.  Although the  net current through the device is zero,
its local  value may be quite large -- in fact larger than the Hall current
density,  defined as the {\it difference} between the total current density
measured in the transport regime and the  current density   at
equilibrium. It is evident  that a clear understanding of the properties of the
{\it equilibrium} current distribution is a prerequisite for the understanding
of  experiments on the Hall current distribution.        In two recent  papers
\cite {Geller1,Geller2} we  studied in detail the  properties of
this distribution, and  found that  they could be summarized in a few
fairly general statements. We considered  a
 confining potential that varies slowly on the scale of the magnetic length
$l$:  $V'l<< \Delta$ where $\Delta$ is the minimum energy gap that we want to
 to take into account ($\Delta = \hbar \omega_c$ if the fractional quantum Hall
effect is ignored).  This condition -- at least insofar as the integral Hall
effect is concerneed -- is  very well satisfied in
 realistic  devices, in which the confinement lengths are of the order of
$10^4 \AA$  or more, while $l$ is of the order of $100$ \AA \cite{Glazman}.
Our expression, which correctly describes the current distribution on a scale
larger than $l$,  has two components:   an ``edge" current,  which is
proportional to the  density gradient, and  a ``bulk" current,  which is
proportional to the gradient of a self-consistent potential.  In the limit of
zero temperature these two contributions become separate in space: The  edge
current flows in the ``compressible" regions, in which the density varies,
while
the self-consistent potential remains constant.  The bulk current  flows
in the ``incompressible" regions,  where the self-consistent potential varies
and the density is constant.  The directions of
these two types of currents generally display a striking alternating pattern.
Furthermore, the integrated currents across a compressible or incompressible
region  are {\it universal}, in the sense of depending only on the  chemical
potentials of a {\it uniform} two-dimensional electron gas near an
incompressible state.

\medskip
{\bf 2. Current Density Distribution}
\medskip

A  convenient  tool for the study of equilibrium orbital currents in an
interacting electronic system is provided by the current-density functional
theory (CDFT) \cite {VR}.  This
formulation provides a rigorous mapping of the many-body problem for the
ground-state energy, density, and current density,  to an effective one-body
problem, in which independent electrons move under the action of
self-consistent local scalar and vector potentials. Unlike the ordinary
density functional theory, CDFT is rigorous, in principle, for the
calculation of equilibrium currents. The effective single particle
equations  have the form  \cite {VR} \begin
{eqnarray}
 ({1 \over 2m} [- i\vec \nabla +{e \over c} \vec A(\vec r)]^2 +V(\vec
r)+ V_H(\vec r)  +V_{xc}(\vec r) + [- i\vec \nabla +{e \over c} \vec
A(\vec r)]\cdot {e \over 2mc} \vec A_{xc}(\vec r) \nonumber \\
 +  {e \over 2mc} \vec A_{xc}(\vec r) \cdot [- i\vec \nabla
+{e \over c} \vec A (\vec r)] -{e \over c} {\vec j(\vec r) \cdot \vec A_{xc}
(\vec r) \over n(\vec r)} ) \psi_\alpha (\vec r) = \epsilon_\alpha \psi_
\alpha (\vec r),
\label {eq21}
\end {eqnarray}
where $\vec A(\vec r)$ and $V(\vec r)$ are the external vector and
scalar potentials, $n(\vec r)$ is the number density, and $\vec j(\vec r)$ is
the
current density. Here
\begin{equation}V_H(\vec r) =  e^2 \int d \vec r' {n(\vec r') \over \vert \vec
r
- \vec r' \vert} \end{equation}
 is the Hartree potential,
\begin{equation}
V_{xc}(\vec r) = {\delta  E_{xc}[n(\vec r),B_\nu(\vec r)] \over \delta n(\vec
r)}  \end{equation}  is the exchange-correlation (xc) scalar potential, and
\begin{equation} {e \over c} \vec A_{xc}(\vec r) =  -{mc \over e n(\vec r)}
\vec
\nabla \times
 {\delta E_{xc}[n(\vec r),B_\nu(\vec r)]  \over \delta \vec B_\nu (\vec r) }
\end{equation}is the exchange-correlation vector potential.  A crucial
feature of the theory is that  the xc energy functional $E_{xc}[n(\vec
r),B_\nu(\vec r)]$ is a functional of the density $n(\vec r)$ and  the {\it
vorticity} \cite {VR} \begin {equation}
\vec B_\nu (\vec r) = \vec B(\vec r) - {mc \over e} \vec \nabla \times
\left ({\vec j (\vec r) \over n(\vec r)} \right ),
\end {equation}
where $\vec B(\vec r) = \vec \nabla \times \vec A (\vec r)$ is the magnetic
field.  The density and orbital  current density are
self-consistently determined by sums of  orbitals,  which are solutions of
eq. (\ref{eq21}). We have ignored spin for simplicity.   The theory  would
yield
the exact density and current distributions, if the exact exchange-correlation
energy functional  were known.
 In the local density approximation (LDA),  which is justified for
density distributions that are slowly varying on the scale of the magnetic
length,   $E_{xc}$ takes the form \begin {equation} E_{xc} [n(\vec r),B_\nu
(\vec
r)]  \simeq \int \epsilon_{xc}[n(\vec r), B_\nu (\vec r)] d \vec r \label
{eq22}
\end {equation}
where $\epsilon_{xc}[n,B]$ is the xc energy density of the two-dimensional
electron gas (2DEG) of uniform density $n$ in a uniform magnetic field $B$.

We now specialize  to the case of a uniform magnetic field perpendicular to the
plane of the electrons  $\vec A(\vec r) = Bx \hat y$ in the Landau gauge,  and
a
slowly varying confining potential $V(\vec r)$.  Following the approach of
\cite{Geller1}, at each point $\vec r_0$ in space we define local cartesian
axes
such that the $x$ axis is along the direction of the gradient of  the total
self-consistent potential $V_{Hxc} \equiv V+V_H+V_{xc}$.  Thus, the total
potential is locally a function of the $x$ coordinate only.  Choosing also the
Landau gauge in such a way that $\vec A$ depends only on the local $x$
coordinate  (and  is parallel to the local $y$ axis)  we effectively obtain a
local one-dimensional problem.  In particular,  the density  and current
profiles depend only on the local $x$ coordinate.  This is so, because an
$x$-dependent density and  current profile leads to an $x$-dependent xc vector
potential in the $y$ direction given by the formula \begin {equation} {e \over
c}
A_{xc} (x) = -{mc \over en(x)} {d \over dx}  {\delta E_{xc} [n(x),B_\nu(x)]
\over \delta B_\nu(x)}.
 \label {eq23}
\end {equation}It is then clear that the
eigenfunctions of the Kohn-Sham equation localized near $\vec r_0$  have the
form
\begin {equation}
 \psi_{nk} (x,y) = {1 \over \sqrt L_y} e^{ i k y} \psi_{nk} (x),
\label {eq24}
\end {equation}
where the $\psi_{nk} (x)$ are  normalized solutions of the equation
\begin {equation}
\left ( - {1 \over 2} {d^2 \over dx^2} + {1 \over 2} (x - X)^2  +  V_{Hxc}(x) +
{e \over c} A_{xc}(x) [x-X-{j_y(x) \over n(x)}] \right ) \psi_{nk} (x) =
\epsilon _{nX} \psi_{nk} (x),
\label {eq25}
\end {equation}
where $x$ is in units of $l$, $n$ is in units of $l^{-2}$,  $j_y$ is
in units of $\hbar /2ml^3$,  $V_{Hxc}$
is in units of $\hbar \omega_c$,  $e A_{xc}/c$ is in units of $\hbar /l$,
and $X=-kl$ is the guiding center of the Landau orbital).  A more rigorous
discussion can be given in terms of exponentially localized Green's functions
\cite{Geller1}.  Eq.~  (\ref{eq25})  may be solved perturbatively,  which is
exact in the limit of slowly varying potential.  Because the eigenfunctions of
$-d^2/dx^2 + (x-X)^2$ are strongly localized about  $x=X$ we can expand \begin
{equation}  V_{Hxc}(x) \approx  V_{Hxc}(X) + (x-X) V_{Hxc}'(X) \label {eq26}
\end {equation}and use
\begin {equation}
{e \over c} A_{xc}(x) [x-X -{j_y(x) \over n(x)}] \approx  {e \over c}
A_{xc}(X).
[x-X] \label {eq27}
\end {equation}
In writing down the second equation we have exploited two facts:
(i)  $A_{xc}(x)$  being given by eq.(\ref{eq23}) is already a gradient of a
local function of $x$.  Hence, we truncate its expansion to the zeroth order in
$x-X$   (ii)  $j_y(x)$,  as it will soon be clear,  is also a gradient of a
local
function of $x$, therefore its multiplication by $A_{xc}(x)$ is the product
of two gradients,  which must be regarded as a higher order infinitesimal
quantity and hence dropped in the present theory.

Using equations (\ref{eq26}) and (\ref{eq27}), equation  (\ref{eq25}) can be
solved, and the density and current density can be computed explicitely.
The general character of the density distribution has been discussed  at length
elsewhere \cite{Glazman,Geller1,Geller2}. Very briefly,  the self-consistent
density profile consists of alternating compressible and incompressible
regions.  A typical profile is shown by the solid line in Fig. 1.  In a
compressible region, the density varies  while the self-consistent potential is
pinned to a constant value.  In an incompressible region,  the density is
pinned
to an integral or fractional multiple of $1/2 \pi l^2$, and the self-consistent
potential varies. The self-consistent chemical potential is tied to a set of
quasi-degenerate single particle orbitals, which are localized in the
compressible regions.  Rather than insisting on the properties of the density
distribution, which are well understood, here  we focus on the result for the
current density, which can be stated as follows: \begin {equation}  j_y(x) =
\gamma_k {dn \over dx}  + 2n(x) [V_{Hxc}'(x) + {e \over c} A_{xc}(x)] \label
{eq28} \end {equation} where  \begin{equation} \gamma_k [n,B]  \equiv (2 [\nu
]+1) ,  \end{equation} and $[\nu]$ is the integral part of the filling factor.
As a final step,  we observe that from eq. ~(\ref{eq23}),  combined with the
local density approximation,  eq. ~(\ref{eq22}),  we obtain  \begin {equation}
2{e \over c} A_{xc}(x) n(x) =   \gamma_{xc} [n,B] {dn (x) \over dx},
 \label {eq29}
\end {equation}
where we have defined
\begin{equation}
\gamma_{xc}[n,B] \equiv {1 \over \mu_B} {\partial \mu_{xc} [n,B] \over \partial
B}.
\end {equation}
Here $\mu_B \equiv e \hbar /2mc$ is the Bohr magneton and
$\mu_{xc} \equiv \partial \epsilon_{xc}[n,B] /\partial n$ is the xc part of the
chemical potential of the uniform 2DEG. We have set $B_\nu = B$ after taking
the
functional derivative of $E_{xc}$, since $A_{xc}$ is already of the desired
order in the gradient of the density.  It is important that, in the definition
of
$\gamma_{xc}[n,B]$, the density is never taken to be exactly equal to one of
the
 incompressible  densities at which $\mu_{xc}$ is discontinuous.  This
prescription assures that $\gamma_{xc}$ is finite everywhere, although possibly
discontinuous at the incompressible densities.

Restoring the physical units and returning to the original coordinate system we
finally write our result as
\begin {equation}
\vec j(\vec r) = {\hbar \over 2m} (\gamma_k[n,B] +\gamma_{xc}[n,B])  \vec
\nabla
n(\vec r) \times \hat z   +  {1 \over m \omega_c} n(\vec r)  \vec \nabla
V_{Hxc}
(\vec r) \times \hat z.
 \label {eq211}
\end {equation}
To understand this formula, we must keep in mind that, in the low temperature
limit, the ``bulk" term ($\vec \nabla V(\vec r)$) and the ``edge" term
 ($\vec \nabla n(\vec r)$) contribute to mutually exclusive regions of space
 -- incompressible and compressible regions, respectively. In the
incompressible regions, the density is  pinned to one of the quantized values,
and the bulk term reduces to the usual Hall current $\vec j =  (e \nu_i /h)
\vec \nabla V_{Hxc} (\vec r) \times \hat z$, where $\nu_i$ is a quantized
filling factor, and $ \vec \nabla V_{Hxc}$ plays the role of the electric
field.
Assuming, for the sake of definiteness,  that the self-consistent potential
increases monotonically from the center to the perifery of a quantum Hall
system
(for example a bar or a dot),  this contribution to the current is  found to be
{\it paramagnetic}.
In the compressible regions, the self-consistent potential is pinned, while the
density varies.  Here the edge term becomes operative.  Under the assumption
that
the density decreases monotonically from the center to the perifery of the
system,  its contribution to the current is found to be {\it diamagnetic}.
A typical self-consistent current density profile is shown as a dashed line in
Fig. 1.

{\bf 3. Orbital Magnetization}

Since the {\it charge} current $\vec J(\vec r) \equiv -e \vec j(\vec r)$
satisfies the continuity equation \begin {equation}
\vec \nabla \cdot \vec J (\vec r) = 0,
 \label {eq31}
\end {equation}
it is possible to write it as the curl of a local function  $M(\vec r) \hat z$
 which we call the ``orbital
magnetization":
\begin {equation}
\vec J(\vec r) = c \vec \nabla \times  M (\vec r) \hat z.
 \label {eq32}
\end {equation}
  It is evident that $M (\vec r)$ is  defined only up to  an
arbitrary additive constant. However, in the limit of slowly varying density,
the orbital magnetization  may be derived from the internal
energy of the uniform electron gas
\begin {equation}
 M(\vec r) =  -  {\partial \epsilon [n(\vec r),B] \over \partial B}.
 \label {eq321}
\end {equation}
This gives \cite{Skudlarski}
\begin {equation}
\vec J(\vec r) = - c {\partial^2 \epsilon [n,B] \over \partial n \partial
B} \vec \nabla n \times \hat z
 \label {eq33}
\end {equation}
with $\epsilon \equiv \epsilon_k + \epsilon_{xc}$ the total internal energy
(kinetic plus exchange-correlation) per unit area of the uniform 2DEG.  It is
not
immediately evident that eq.(\ref{eq33}) is equivalent to eq.(\ref{eq211}).
A puzzling feature is that the $\vec \nabla V (\vec r)$ term does not appear
explicitly in eq.(\ref{eq33}).  The puzzle is resolved by noting
that the thermodynamic derivative appearing in eq.~(\ref{eq33})  has
singularities when the density is such that the compressibility vanishes.
These
singular contributions are responsible for the appearance of the
$\vec \nabla V (\vec r)$ term in the incompressible regions. To show
this,  we rewrite $\partial^2 \epsilon /\partial n \partial B$ as \begin
{equation}  {\partial^2 \epsilon [n,B] \over \partial n \partial
B} = \left ({\partial \mu \over \partial B}\right )_n = - {(\partial n /
\partial
B)_\mu \over (\partial n / \partial \mu)_B}.
 \label {eq34}
\end {equation}
When the density  is {\it not} such that the compressibility  vanishes, the
derivatives are finite, and it is easy to verify that eq.~ (\ref{eq33})
reduces  to the first term of eq.~(\ref{eq211}).   But when the density is such
that the compressibility vanishes, we must turn to the second equality in
(\ref{eq34}), for insight into the behavior of the current. The denominator
$(\partial n / \partial \mu)_B$ vanishes, because the system is incompressible.
At the same time $\vec \nabla n(\vec r)$ also vanishes in the regions where
incompressibility holds.  The ratio of these two quantities, however,  remains
finite in the limit $T \to 0$: \begin{equation}
- \left ( {\partial \mu \over \partial n} \right )_B \vec \nabla n(\vec r) \to
\vec \nabla V_{Hxc}(\vec r).
\end {equation}
The numerator $(\partial n / \partial B)_\mu$  is well known to be proportional
to the Hall conductivity $e^2 \nu_i /h$ \cite{Streda}.  This establishes the
equivalence of  eqs.~ (\ref{eq33})  and (\ref{eq211}).

The usefulness of writing the current as the curl of the orbital magnetization
lies in the fact that it  makes possible to express the flux of current across
an
arbitrary path as the difference in the values of the orbital magnetization at
the end points: \begin {equation}
I_{1,2} =  \int_1^2 \vec j (\vec r) \cdot \hat n (\vec r) dl = c[M(1)-M(2)]
\label {eq15},
\end{equation}
where the line integral is along a path joining  points $1$ and $2$, and
$\hat n$ is the unit vector normal to the path.
  The key observation is that the orbital magnetization at the end points may
be
well represented by the LDA even if it is not well represented at intermediate
points  along the path.

As a simple example,  consider the disk of non-interacting electrons studied
recently by Avishai and Kohmoto \cite{Avishai} (AK).  The integrated current
across a radius is simply  the magnetization at the center of the dot,
since the magnetization obviously vanishes at the edge of the system.  But the
density is uniform at the center of the dot,  so the LDA is certainly valid
there.  Furthermore,  under the conditions envisaged by AK,  the Fermi energy
is
pinned to the bottom of the  i-th Landau level in the bulk,  which means  that
$\mu = (i+1/2) \hbar \omega_c^+$  (a  superscript ``+" or ``--" means that an
infinitesimal positive quantity must be added or subtracted).  Looking at the
relation between $n$ and $\mu$ in the non-interacting 2DEG at an infinitesimal
temperature we see that this chemical potential corresponds to a
density $n = [(i+1)/2 \pi l^2]^-$ at the center of the dot.  This, in turn,
corresponds to an orbital  magnetization $M = -(i \mu_B/2 \pi l^2)$, as one
can see from Fig. 2.   Therefore the total current integrated from the center
to
the edge of the dot is equal to $-i e \omega_c / 4 \pi$, i.e., the  total
current is quantized in integral multiples of $e \omega_c /4 \pi$.  This result
was   first  derived  by AK from a numerical solution of the Schr\"odinger
equation with hard wall boundary conditions. The present derivation
demonstrates the generality of the  result.

Next we consider the interacting electron fluid. The energy of the uniform
phase consists of kinetic and exchange-correlation contributions. The kinetic
energy is known exactly, and the  kinetic contribution to the orbital
magnetization is given by
\begin{equation}
M_k(\nu,B) = \left ( [\nu] - (2 [\nu] +1)(\nu - [\nu] \right ) {\mu_B \over 2
\pi
l^2}. \end {equation}
This curve is plotted in Fig. 2.  No
simple expression is known which represents $\epsilon_{xc}$ accurately at all
values of filling factor $\nu$.  However, to calculate the integrated current
in
a compressible or incompressible region, we only need to know what happens in
the vicinity of an incompressible filling factor $\nu_0$, integral or
fractional.  Here some essentially
 exact results can be obtained.  In a neighborhood of $\nu_0$ the
exchange-correlation   energy {\it density} can be expanded as follows:
\begin {equation}
\epsilon_{xc}(\nu,B) = \left [\epsilon_{xc0}   + {(\nu - \nu_0) \over 2 \pi
l^2}
\mu_{xc}[\nu_0^+,B] +O((\nu - \nu_0)^{3/2}) \right ]
\label {eq16} \end{equation} for $\nu > \nu_0$,  and
\begin {equation}
\epsilon_{xc} (\nu,B) = \left [\epsilon_{xc0} + {(\nu - \nu_0) \over 2 \pi l^2}
\mu_{xc}[\nu_0^-,B] +O((\nu_0 - \nu)^{3/2}) \right ]
\label {eq17}
\end{equation}
for $\nu < \nu_0$.
 Here $\epsilon_{xc0}$  is the xc energy density of the incompressible state at
filling factor $\nu_0$  and
$\mu_{xc}[\nu_0^+,B]$ and  $\mu_{xc}[\nu_0^-,B]$ are the   xc chemical
potentials calculated as right or left derivatives of the xc energy density
with respect to density.  In the limit of high magnetic field ($e^2/l \hbar
\omega_c <<1$) $\epsilon_{xc0}$ scales as $\epsilon_{xc0} (B) = \bar
\epsilon_{xc0} e^2 / 2 \pi l^3 $ and the $\mu_{xc} $'s scale as
$\mu_{xc}(\nu,B) = \bar \mu_{xc} (\nu) e^2/l$, where the dimensionless
quantities
$\bar \epsilon_{xc0}$ and $\bar \mu_{xc} (\nu) $ are independent of
magnetic field.
 The $O((\nu - \nu_0)^{3/2})$ term on the right hand side
represents the interaction energy of a classical Wigner crystal of dilute
quasiparticles  (or quasiholes) added to the incompressible state.   It is easy
to verify that this term -- as well as all the higher order terms in the
expansion of the energy in $\nu - \nu_0$ -- does not contribute to the orbital
magnetization in the limit $\nu \to \nu_0^{\pm}$.  Taking the derivatives of
eq.
(\ref{eq16}) and   eq. (\ref{eq17}) with respect to $B$, using the power law
scaling of the energy and chemical potentials as functions of $B$,  we obtain
the exchange-correlation contribution to the magnetization
 \begin {equation}
 M_{xc} (\nu = \nu_o^{\pm},B) = \left  [- 3  \bar \epsilon_{xc0} + 2
\nu_0  \bar \mu_{xc}(\nu_0^{\pm}) \right ] {e^2 \over l \hbar \omega_c}  {\mu_B
\over 2 \pi l^2}
\label {eq18} \end{equation}

The above equation may be used to calculate the exact total current that flows
through a compressible strip or channel connecting an incompressible region at
filling factor $\nu_0$ to another incompressible region at filling factor
$\nu_1$.  Assuming,  for definiteness, that $\nu_0 > \nu_1$ and
integrating the current flow along a line going from the higher to the lower
density we find that the total current is
\begin {equation}
I = c[M(\nu=\nu_0^-,B)-M(\nu=\nu_1^+,B)],
\end {equation}
where $M=M_k+M_{xc}$. The remarkable feature of this result is that it holds
true despite the lack of detailed information about the density and current
distribution within the compressible region, i.e., despite the fact that
$n(\vec
r)$ and $\gamma_{xc}(n,B)$ are not known everywhere inside the region.
Equation (\ref{eq18}) can also be used to calculate the
current flowing through the incompressible region,  say at filling factor
$\nu_1$ .  For this region we obtain
\begin{equation}
I = c[M(\nu_1^+) -  M(\nu_1^-)]
= {e \nu_1 \over h} \Delta_1,
\end {equation}
where $\Delta_1 = \mu(\nu_1^+,B) - \mu(\nu_1^-,B)$ is the  gap in the electron
chemical potential at filling factor $\nu_1$.  This result is precisely what
one would expect, based on the observation that the change in the
self-consistent potential across an incompressible strip must equal the
incompressibility gap.

{\bf 4. Conclusions}

In conclusion, we have elucidated in this paper the nature of the equilibrium
current distribution in a two-dimensional electron gas confined by a potential
which varies  slowly on the scale of the magnetic length.  We have
shown that the density profile consists of alternating compressible and
incompressible regions,  and that the current flows in opposite directions in
these two types of regions,  reflecting the sawtooth behavior of the
magnetization  of the uniform interacting electron gas as a function of
density.  Finally,  we have demonstrated that the integrated current across a
compressible or incompressible region can be expressed (in the limit of strong
magnetic field)  in terms of the chemical potentials  of the uniform electron
gas
near  incompressible filling factors.  Detailed numerical results for
exchange-correlation chemical potentials near integral filling factors will be
reported elesewhere.

{\bf Acknowledgements}
We acknowledge support from NSF grant No. DMR-9403908.

\newpage

\begin{figure}
\caption{Typical  self-consistent equilibrium density (solid curve) and current
density (dashed curve) for a two dimensional electron fluid confined along the
$x$ direction at low temperature.  The density is plotted in units of $\rho_0
= 1/2 \pi l^2$ and the current is plotted in units of $j_0 = e \omega_c/2 \pi
l$.  Note the alternating signs of the edge and bulk currents}
 \label{fig1}
\end{figure}

\begin{figure}
\caption{Orbital magnetization density versus filling factor for a
non-interacting uniform two-dimensional electron gas at low temperature.}
 \label{fig2}
\end{figure}
\end{document}